\documentclass[10pt,conference,english]{IEEEtran}
\usepackage[centertags]{amsmath}
\usepackage{amsfonts}
\usepackage{amssymb}
\usepackage{epsfig}
\usepackage{amsmath,  amssymb}
\usepackage{graphicx}
\usepackage[centertags]{amsmath}
\usepackage{clrscode}
\usepackage{cite}
\usepackage{fullpage}
\usepackage{amsmath,amssymb,mathrsfs,pseudocode}
\usepackage{color}
\usepackage[normalem]{ulem}
\usepackage{psfrag}
\usepackage{url}
\usepackage{babel}

\usepackage[margin=18.5mm]{geometry}
\usepackage{xspace}
\usepackage{multicol}
\usepackage{booktabs} 
\usepackage{multirow}
\usepackage[x11names,dvipsnames,table]{xcolor}

\newcommand{\myhline}{\hline}
\newcommand{\opertype}[1]{\begin{minipage}{29mm}\centering\vspace{1mm} #1\vspace{1mm}\end{minipage}}

\newtheorem{theorem}{Theorem}

\newtheorem{example}[theorem]{Example}

\usepackage{xspace}
\usepackage{bbm}
\catcode`@=11
\chardef\mathlig@atcode\count255

\def\actively#1#2{\begingroup\uccode`\~=`#2\relax\uppercase{\endgroup#1~}}
\def\mathlig@gobble{\afterassignment\mathlig@next@cmd\let\mathlig@next= }
\def\mathlig@delim{\mathlig@delim}
\def\mathlig@defcs#1{\expandafter\def\csname#1\endcsname}
\def\mathlig@let@cs#1#2{\expandafter\let\expandafter#1\csname#2\endcsname}
\def\mathlig@appendcs#1#2{\expandafter\edef\csname#1\endcsname{\csname#1\endcsname#2}}

\def\mathlig#1#2{\mathlig@checklig#1\mathlig@end\mathlig@defcs{mathlig@back@#1}{#2}\ignorespaces}

\def\mathlig@checklig#1#2\mathlig@end{%
 \expandafter\ifx\csname mathlig@forw@#1\endcsname\relax
 \expandafter\mathchardef\csname mathlig@back@#1\endcsname=\mathcode`#1%
 \mathcode`#1"8000\actively\def#1{\csname mathlig@look@#1\endcsname}%
 \mathlig@dolig#1\mathlig@delim
\fi
\mathlig@checksuffix#1#2\mathlig@end
}

\def\mathlig@checksuffix#1#2\mathlig@end{%
\ifx\mathlig@delim#2\mathlig@delim\relax\else\mathlig@checksuffix@{#1}#2\mathlig@end\fi
}
\def\mathlig@checksuffix@#1#2#3\mathlig@end{%
\expandafter\ifx\csname mathlig@forw@#1#2\endcsname\relax\mathlig@dosuffix{#1}{#2}\fi
\mathlig@checksuffix{#1#2}#3\mathlig@end
}

\def\mathlig@dosuffix#1#2{%
\mathlig@appendcs{mathlig@toks@#1}{#2}%
\mathlig@dolig{#1}{#2}\mathlig@delim
}

\def\mathlig@dolig#1#2\mathlig@delim{%
 \mathlig@defcs{mathlig@look@#1#2}{%
 \mathlig@let@cs\mathlig@next{mathlig@forw@#1#2}\futurelet\mathlig@next@tok\mathlig@next}%
 \mathlig@defcs{mathlig@forw@#1#2}{%
  \mathlig@let@cs\mathlig@next{mathlig@back@#1#2}%
  \mathlig@let@cs\checker{mathlig@chck@#1#2}%
  \mathlig@let@cs\mathligtoks{mathlig@toks@#1#2}%
  \expandafter\ifx\expandafter\mathlig@delim\mathligtoks\mathlig@delim\relax\else
  \expandafter\checker\mathligtoks\mathlig@delim\fi
  \mathlig@next
 }%
 \mathlig@defcs{mathlig@toks@#1#2}{}%
 \mathlig@defcs{mathlig@chck@#1#2}##1##2\mathlig@delim{%
  \ifx\mathlig@next@tok##1%
   \mathlig@let@cs\mathlig@next@cmd{mathlig@look@#1#2##1}\let\mathlig@next\mathlig@gobble
  \fi 
  \ifx\mathlig@delim##2\mathlig@delim\relax\else
   \csname mathlig@chck@#1#2\endcsname##2\mathlig@delim
  \fi
 }%
 \ifx\mathlig@delim#2\mathlig@delim\else
  \mathlig@defcs{mathlig@back@#1#2}{\csname mathlig@back@#1\endcsname #2}%
 \fi
}%

\catcode`@\mathlig@atcode

\newcommand{\muspace}{\mspace{1mu}}

\DeclareRobustCommand{\scond}{\mathchoice{\muspace\vert\muspace}{\vert}{\vert}{\vert}}
\mathlig{|}{\scond}

\DeclareRobustCommand{\discint}{\mathchoice{\mspace{-1.5mu}:\mspace{-1.5mu}}{\mspace{-1.5mu}:\mspace{-1.5mu}}{:}{:}}
\mathlig{::}{\discint}
\newcommand{\suchthat}{\colon}

\newcommand{\Cr}{\mathscr{C}}

\newcommand{\Rr}{\mathscr{R}}

\newcommand{\Cv}{{\bf C}}
\newcommand{\Dv}{{\bf D}}

\newcommand{\Rv}{{\bf R}}
\newcommand{\Sv}{{\bf S}}

\newcommand{\tv}{{\bf t}}

\newcommand{\rv}{{\bf r}}

\newcommand{\Xh}{{\hat{X}}}

\newcommand{\xh}{{\hat{x}}}

\def\b{\beta}

\let\P\relax
\DeclareMathOperator\P{\textsf{P}}

\DeclareMathOperator\R{\textsf{R}}

\def\textiid{i.i.d.\@\xspace}
\newcommand\iid{\ifmmode\text{ i.i.d. } \else \textiid \fi}

\def\clap#1{\hbox to 0pt{\hss#1\hss}}
\def\mathclap{\mathpalette\mathclapinternal}
\def\mathclapinternal#1#2{%
  \clap{$\mathsurround=0pt#1{#2}$}}

\let\oldstackrel\stackrel
\renewcommand{\stackrel}[2]{\oldstackrel{\mathclap{#1}}{#2}}

\begin{document}
\title{On the Capacity for Distributed Index Coding\vspace{-2mm}}

\author{
  Yucheng Liu, Parastoo Sadeghi\\      Research School of Engineering\\ Australian National University\\ \{yucheng.liu, parastoo.sadeghi\}@anu.edu.au
  \and
  Fatemeh Arbabjolfaei, Young-Han Kim\\      Department of Electrical and Computer Engineering\\ University of California, San Diego\\ \{farbabjo, yhk\}@ucsd.edu
}


\maketitle

\begin{abstract}
The distributed index coding problem is studied, whereby multiple messages are stored at different servers to be broadcast to receivers with side information. First, the existing composite coding scheme is enhanced for the centralized (single-server) index coding problem,
which is then merged with fractional partitioning of servers to yield a new coding scheme for distributed index coding.
New outer bounds on the capacity region are also established. For 213 out of 218 non-isomorphic distributed index coding problems with four messages the achievable sum-rate of the proposed distributed composite coding scheme matches the outer bound, thus establishing the sum-capacity for these problems.
\end{abstract}

\vspace{-1mm}
\section{Introduction}\label{sec:intro}

Introduced by Birk and Kol \cite{Birk--Kol1998}, the index coding problem studies the optimal broadcast rate from a server to multiple receivers with some side information about the messages.
This paper considers the distributed index coding problem in which, unlike the aforementioned single-server index coding problem, the messages are distributed over multiple servers. 
The distributed index coding problem was first studied by Ong, Ho, and Lim \cite{Ong--Ho--Lim2016}, where lower and upper bounds on the (optimal) broadcast rate were derived in the special case in which each receiver has a distinct message as side information and it is shown that the bounds match if no two servers have any messages in common.
Thapa, Ong, and Johnson \cite{Thapa--Ong--Johnson2016} considered the distributed index coding problem with two servers each having an arbitrary subset of messages and extended some of the existing schemes for the centralized index coding to the two-server distributed case.

The main objective of this paper is to study a general distributed index coding problem and to establish new inner and outer bounds on the capacity region. 
For the inner bounds, we propose distributed composite coding schemes that extend those in \cite{Sadeghi--Arbabjolfaei--Kim2016} through an enhanced version of the centralized composite coding scheme.
The enhancement is interesting on its own and strictly improves upon  the original composite coding scheme in \cite{Arbabjolfaei--Bandemer--Kim--Sasoglu--Wang2013}.
For the outer bounds, we establish a set of new inequalities to add to the existing polymatroidal outer bound derived in \cite{Sadeghi--Arbabjolfaei--Kim2016}. The inner and outer bounds, when applied to the sum-rate, match in 213 out of 218 non-isomorphic distributed index coding problems with four messages, thus establishing their sum-capacity. Note that in \cite{Sadeghi--Arbabjolfaei--Kim2016}, the capacity region for all three-message distributed index coding problems was established.

Throughout the paper, $[n]$ denotes the set $\{1, 2, \ldots, n\}$ and
$N \doteq \{J \subseteq [n] \suchthat J \not = \emptyset\}$ denotes the set of all nonempty subsets of $[n]$.
Given a tuple $(x_1, \ldots, x_n)$ and $A \subseteq [n]$, $x(A)$ denotes the subtuple $(x_i: i \in A)$.

\section{System Model and Problem Setup}\label{sec:model}

\begin{figure}
\begin{center}
\small
\psfrag{x1}[b]{$x_1$}
\psfrag{x2}[b]{$x_2$}
\psfrag{x3}[b]{$x_3$}
\psfrag{x12}[b]{$x_1, x_2$}
\psfrag{x13}[b]{$x_1, x_3$}
\psfrag{x23}[b]{$x_2, x_3$}
\psfrag{x123}[b]{$x_1, x_2, x_3$}
\psfrag{y1}[b]{$y_{\{1\}}$}
\psfrag{y2}[b]{$y_{\{2\}}$}
\psfrag{y3}[b]{$y_{\{3\}}$}
\psfrag{y12}[b]{$y_{\{1,2\}}$}
\psfrag{y13}[b]{$y_{\{1,3\}}$}
\psfrag{y23}[b]{$y_{\{2,3\}}$}
\psfrag{y123}[b]{$y_{\{1,2,3\}}$}
\psfrag{e1}[c]{Server $\{1\}$}
\psfrag{e2}[c]{Server $\{2\}$}
\psfrag{e3}[c]{Server $\{3\}$}
\psfrag{e12}[c]{Server $\{1,2\}$}
\psfrag{e13}[c]{Server $\{1,3\}$}
\psfrag{e23}[c]{Server $\{2,3\}$}
\psfrag{e123}[c]{Server $\{1,2,3\}$}
\psfrag{d1}[c]{Decoder $1$}
\psfrag{d2}[c]{Decoder $2$}
\psfrag{d3}[c]{Decoder $3$}
\psfrag{xh1}[b]{$\xh_1$}
\psfrag{xh2}[b]{$\xh_2$}
\psfrag{xh3}[b]{$\xh_3$}
\psfrag{a1}[b]{$x({A_1})$}
\psfrag{a2}[b]{$x({A_2})$}
\psfrag{a3}[b]{$x({A_3})$}
\includegraphics[scale=0.34]{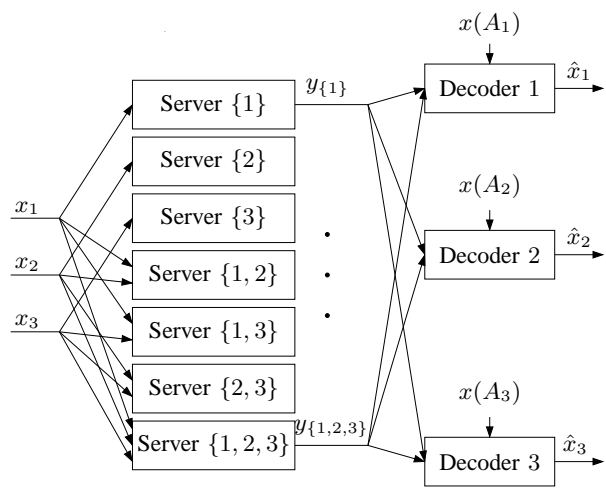}
\end{center}
\caption{The distributed index coding problem with $n=3$.}\vspace{-5mm}
\label{fig:index_coding}
\end{figure}

Consider the general distributed index coding problem in which there are $n$ messages $x_1, x_2, \ldots, x_n$, where $x_i \in \{0,1\}^{t_i}$, $i \in [n]$.
As shown in Fig.~\ref{fig:index_coding}, there are $2^n-1$ servers, where server $J \in N$ has access to messages $x(J)$.
There are $n$ receivers, where receiver $i \in [n]$ wishes to obtain message $x_i$ and knows $x(A_i)$ as side information for some $A_i \subseteq [n]\setminus \{i\}$.
Server $J$ is connected to all receivers via a noiseless broadcast link of finite capacity $C_J$. 
This is a fairly general model that allows for all possible message availabilities on different servers. 
If $C_J = 1$ only for $J= [n]$ and is zero everywhere else, we recover the centralized index coding problem. 
The question is to find the maximum amount of information that can be communicated to the receivers and the optimal coding scheme that achieves this maximum.
To answer this question formally, we define 
a $(\tv,\rv) = ((t_i, i \in [n]), (r_J, J \in N))$ {\em distributed index code} by
\begin{itemize}
\item $2^n-1$ encoders, one for each server $J \in N$, such that $\phi_J: \prod_{j \in J} \{0,1\}^{t_j} \to \{0,1\}^{r_J}$ maps the messages in server $J$, $x(J)$, to an $r_J$-bit sequence $y_J$, and
\item $n$ decoders $\psi_i: \prod_{J \in N} \{0,1\}^{r_J} \times \prod_{k \in A_i} \{0,1\}^{t_k} \to \{0,1\}^{t_i}$ that maps the received sequences $\phi_J(x_j, j \in J)$ and the side information $x(A_i)$ to 
$\hat{x}_i$ for $i \in [n]$.
\end{itemize}

Let $X_i$ and $\Xh_i$ be the random variables representing message $x_i$ and message estimate $\xh_i$, respectively.
Assume that $X_1, \ldots ,X_n$ are uniformly distributed and independent of each other.
We say that a rate--capacity tuple $(\mathbf{R}, \mathbf{C}) = ((R_i, i \in [n]),(C_J, J \in N))$ is achievable if for every $\epsilon > 0$, there exist a $(\tv,\rv)$ code and $r$ such that
\[
R_i \le \frac{t_i}{r}, \quad i \in [n], \qquad \qquad
C_J \geq \frac{r_J}{r}, \quad J \in N,
\]
and the probability of error 
\[
\P\{ (\Xh_1,\ldots, \Xh_n) \ne (X_1, \ldots, X_n)\} \le \epsilon.
\]
For a given $\mathbf{C} = (C_J, J \in N)$, the capacity region $\Cr$ of this index coding problem is the closure of the set of all $\Rv=(R_1, \cdots, R_n)$ such that $(\Rv,\Cv)$ is achievable.
Although the capacity region of the centralized index coding problem is known to be equal to its zero-error capacity region \cite{Langberg--Effros2011}, it is not known whether these two capacity regions are equal for the distributed case.
Throughout the paper, we will compactly represent a distributed index coding instance (for a given $\mathbf{C}$) by a sequence  $(i|j \in A_i)$, $i \in [n]$. For example, for $A_1 = \emptyset$, $A_2 = \{3\}$, and $A_3=\{2\}$, we write
$(1|-),\, (2|3),\, (3|2)$.

\section{Composite Coding for the Centralized Case}\label{sec:centralized:bounds}
\def\R{\mathcal R}
\newcommand{\ckj}{S_{K,J}}
\newcommand{\ck}{S_{K}}
\newcommand{\ckd}{S_{K}(\Dv)}
\newcommand{\proj}{\text{Proj}}
\newcommand{\CO}{\text{co}}
\newcommand{\D}{\mathcal{D}}
\newcommand{\ckjd}{S_{K,J}(\Dv)}
\newcommand{\ckjpdp}{S_{K,J}(P,\Dv)}
\newcommand{\rcc}{\Rr_\mathrm{CC}}
\newcommand{\rcce}{\Rr_\mathrm{CC}^{(e)}}
\newcommand{\bcc}{\b_\mathrm{CC}}
\newcommand{\bcce}{\b_\mathrm{CC}^{(e)}}

We present a new composite coding scheme for the centralized index coding problem, which is an extension of the original composite coding scheme proposed in \cite{Arbabjolfaei--Bandemer--Kim--Sasoglu--Wang2013}.

\subsection{Existing Centralized Composite Coding Scheme}\label{sec:existing:centralized}

First, we briefly review the scheme of \cite{Arbabjolfaei--Bandemer--Kim--Sasoglu--Wang2013}. 
Our slightly modified presentation here will lead to better understanding of both centralized and distributed composite coding schemes that are developed in this paper.
There is a single server containing all messages in $[n]$, which is connected to the receivers via a noiseless broadcast channel of capacity $C$. 
To each non-empty subset $K \subseteq [n]$ (or $K \in N)$ of the messages, we associate a virtual encoder with composite coding rate $\ck$. 
In the first step of composite coding, the virtual encoder $K$ maps $x(K)$ into a single composite index $w_{K}$, which is generated randomly and independently as a Bern(1/2) sequence of length ${r \ck}$ bits. 
In the second step, the server uses flat coding \cite{Arbabjolfaei--Bandemer--Kim--Sasoglu--Wang2013} to encode the composite indices $(w_{K}, K \in N)$ into a single sequence $y \in \{0,1\}^{r}$. 

As with the encoding, decoding also takes place in two steps. 
Each receiver first uses its side information to recover all composite indices $(w_{K}, K \in N)$. 
This is successful with vanishing probability of error (for $\tv, \rv$, and $r$ sufficiently large) if 
  \begin{align}\label{eq:flat:centralized}\sum_{K:K \not \subseteq A_{i}} \ck < C, \quad  i \in [n].\end{align}  
As the second step of decoding, each receiver recovers the desired message (together with a subset of other messages) from the composite
indices and the side information. 
Let $D_i$ be the set of the messages that receiver $i$ recovers (due to index coding requirements $i\in D_{i}$) and $\Delta = \D_1 \times \cdots \times \D_n$ be the set of all possible decoding set tuples across all receivers, where $\D_i = \{D_{i}|D_i\subseteq [n]\setminus A_i:i \in D_{i}\}$ is the set of all possible decoding sets for receiver $i$. 
Assume $\Dv \in \Delta$ is the chosen decoding set tuple.
Then, receiver $i$ can successfully recover messages in $D_i$ with vanishing probability of error if
\begin{align}\label{eq:index:coding:centralized}
\sum_{j \in \textcolor{black}{L}} R_j <\sum_{K\subseteq (D_{i}\cup A_{i}):\,K\cap L \neq \emptyset} \ck,
\end{align}
for all $L \subseteq D_{i}$. 
Now let $(\mathbf{R}(\Dv), \mathbf{S})$ be the set of index coding rate tuples and composite rate tuples that satisfy \eqref{eq:flat:centralized} and \eqref{eq:index:coding:centralized} for a given decoding choice $\Dv \in \Delta$, where $\mathbf{R}(\Dv) = (R_i, i\in [n])$ and $\mathbf S = (\ck, K \in N)$. 
The overall achievable rate region $\rcc$ can be written as  
   \begin{align}\label{eq:method1}
\rcc = \proj{ \left[\CO \left[\bigcup_{\Dv \in \Delta}  (\mathbf{R}(\Dv), \mathbf{S})\right]\right]},
\end{align}
 where ``$\CO$'' denotes the convex hull and 
``$\proj$'' 
denotes projecting $(\Rv,\Sv)$ into $\Rv$ coordinates.
It can be shown that $\rcc$ can equivalently be computed as 
   \begin{align}\label{eq:method2}
\rcc = \CO{ \left[\bigcup_{\Dv \in \Delta} \proj(\mathbf{R}(\Dv), \mathbf{S})\right]}.
\end{align}

\subsection{Enhanced Composite Coding}

The main idea behind this new method is to allow the composite coding rates $\ck$ to 
depend on the decoding choices of the receivers. 
Effectively,  composite coding rates can be individually tailored to different decoding choices, as long as they collectively satisfy the conditions of \eqref{eq:flat:centralized}.  
Splitting the rate of each message, we represent the message $x_i$ by independent parts $x_i(\Dv)$ at rate $R_i(\Dv)$.
Thus, 
\begin{align}\label{eq:sumrate:method3:centralized}
R_i = \sum_{\Dv \in \Delta} R_{i}(\Dv), \quad i \in [n].
\end{align}
To send $(x_i(\Dv), i \in [n])$, we generate composite messages $w_K(\Dv)$  at composite coding rate $\ckd$, $K \in N$.
Each receiver first recovers all composite
indices $(w_K(\Dv), K \in N, \Dv \in \Delta)$. 
This is successful with vanishing probability of error if
  \begin{align}\label{eq:flat:method3:centralized}
  \sum_{\Dv \in \Delta}\sum_{K:K \not \subseteq A_{i}} \ckd < C, \quad  i \in [n].
 \end{align}
For a given $\Dv \in \Delta$, 
receiver $i$ can successfully recover messages in $D_i$ 
with vanishing probability of error if

\begin{align}\label{eq:index:coding:method3:centralized}
\sum_{j \in \textcolor{black}{L}} R_j(\Dv) <\sum_{K\subseteq (D_i\cup A_{i}):\,K\cap L \neq \emptyset} \ckd,
\end{align}
for all $L \subseteq D_i$. 
The enhanced achievable rate region $\rcce$ is obtained by projecting out $(\ckd, K \in N, \Dv \in \Delta)$ and $(R_i(\Dv), i \in [n], \Dv \in \Delta)$ through Fourier-Motzkin elimination \cite[Appendix D]{elgamal_yhk}.
The enhanced composite coding inner bound is no smaller than the original composite coding inner bound due to a larger degrees of freedom in choosing the composite rates for different decoding choices $\Dv$.
The enhanced inner bound can also be strictly larger than the original one, as illustrated in the following.
\begin{example}
For index coding problem 
\begin{gather*}
(1|3,4), (2|4,5),\, (3|5,6),\, (4|2,3,6), (5|1,4,6),\, (6|1,2),
\end{gather*}
the largest symmetric rate in $\rcc$ is $0.2963$, while the largest symmetric rate in $\rcce$ is  $0.2987$, and thus $\rcc \subsetneq \rcce$.
\end{example}
A possible disadvantage of this method is its computational complexity due to the increase in the number of composite coding rate variables and the necessity to perform a single Fourier-Motzkin elimination operation on all $(\ckd)$ and $(R_i(\Dv))$.  
To overcome this, one can either apply the technique to a subset of decoding choices in $\Delta$ (possibly at the expense of some reduction in the rate region)  or use linear programming (LP) to solve for a desired weighted sum-rate or equal rates subject to \eqref{eq:sumrate:method3:centralized}-\eqref{eq:index:coding:method3:centralized}. 

\section{Composite Coding for the Distributed Case}\label{sec:distributed:bounds}

In this section, we present new composite coding schemes for the distributed index coding problem. 
First, we apply the enhanced composite coding to all servers as a single group, which is an extension of \cite[Section IV-B]{Sadeghi--Arbabjolfaei--Kim2016}.
Next, we apply the enhanced composite coding to fractional partitions of the servers, which is an extension of the partitioned distributed composite coding \cite[Section IV-C]{Sadeghi--Arbabjolfaei--Kim2016}.

\subsection{All-server Distributed Composite Coding}\label{sec:distributed:bounds:all:servers}

Using rate splitting, similar to the centralized case, we have 
  \begin{align}\label{eq:sumrate:method3:dist}
   R_i = \sum_{\Dv \in \Delta} R_{i}(\Dv), \quad  i \in [n].\end{align}
For each non-empty subset $K \subseteq J$ at server $J \in N$ and each decoding choice $\Dv \in \Delta$, there is a virtual encoder at server $J$.
In the first step of composite coding,  virtual encoder $K$ at server $J$ maps $(x_i(\Dv), i \in K)$ into a composite index $w_{K,J}(\Dv)$ with rate $\ckjd$, which is generated randomly and independently as a Bern(1/2) sequence of length ${r_J \ckjd}$ bits. 
In the second step, server $J$ uses flat coding to encode the composite indices $(w_{K,J}(\Dv), K \subseteq J, \Dv \in \Delta)$ into $y_J \in \{0,1\}^{r_J}$. 

Each receiver first recovers all composite indices $(w_{K,J}(\Dv))$, which is successful with vanishing probability of error if
\begin{align}\label{eq:method3:dist2}
 \sum_{\Dv \in \Delta}\sum_{K:K \nsubseteq A_{i}} \ckjd \le C_J, \quad  i \in [n],  J \in N.
  \end{align}
As the second step of decoding, for each decoding choice $\Dv \in \Delta$, receiver $i \in [n]$ recovers $(x_j(\Dv), j \in D_i)$ (which includes the desired message $x_i(\Dv)$) using the composite indices and its side information.
This is successful with vanishing probability of error if 
\begin{align}\label{eq:method3:dist1}
\sum_{j \in L} R_{j}(\Dv) <\sum_{K\subseteq (D_i\cup A_{i}):\,K\cap L \neq \emptyset} \,\,\sum_{J: K\subseteq J}\ckjd,
\end{align}
for all $L \subseteq D_i$. 
The second summation on the right hand side of the above ensures that all servers that contain the message subset $K$ are taken into account. 
  
The computational complexity of the enhanced composite coding for distributed index coding is even higher than its centralized counterpart, since the number of composite coding rates rapidly grows with the number of servers. 
For each decoding set tuple $\Dv \in \Delta$, there are $\sum_{k=1}^n{n \choose k} (2^k-1)$ composite coding rates $\ckjd$ and $n$ message rates $R_{i}(\Dv)$.
Hence, even for $n=4$ and $|\Delta| = 2$, the number of variables  to eliminate is $2\times(65+4)$.
However, an LP can be solved subject to \eqref{eq:sumrate:method3:dist}-\eqref{eq:method3:dist1}.

Note that in the scheme proposed in \cite[Section IV-B]{Sadeghi--Arbabjolfaei--Kim2016}, the rates of the composite messages $\ckj, K \subseteq J, J \in N$ did not depend on the choice of the decoding set tuple $\Dv$.
%

\subsection{Fractional Distributed Composite Coding}\label{sec:distributed:bounds:fractional}

In the previous scheme, all servers participated in a single group to perform composite coding. However, it is also possible that servers form different groups and participate in group-based composite coding. For each group, they allocate a fraction of their server capacity, hence the name ``fractional''. 

Let $\Pi$ be the collection of non-empty subsets of $N$. 
For each $P \in \Pi$, let $I(P) = \{i| \exists J \in P: i\in J\} \subseteq [n]$ be the union of all messages held by at least one server in $P$. 
For every $i \in I(P)$, let $A_{i}(P) = A_i \cap I(P)$,
$\D_{i}(P) = \{D_i(P)|D_i(P)\subseteq I(P)\setminus A_i(P):i \in D_i(P)\}$,
and $\Delta(P) = \D_{1}(P) \times \cdots \times \D_{n}(P)$.

Essentially, we now apply the all-server distributed composite coding methodology to each server group. 
Then, each receiver is able to decode its desired message with vanishing probability of error if the following sets of inequalities are satisfied.
For all $P \in \Pi$ and $J \in P$
            \begin{align}
            \label{eq:1}
            \sum_{\Dv \in \Delta(P)}\,\,\sum_{K:K \not \subseteq A_i(P)} \ckjpdp \le C_{J}(P),  \quad i \in I(P).
            \end{align}
For all $P \in \Pi$, $i \in I(P)$ and a given $\Dv \in \Delta (P)$, \eqref{eq:method3:dist1} is modified as    
\begin{align}
\sum_{j \in \textcolor{black}{L}} R_{j}(P, \Dv) <\sum_{K\subseteq (D_i\cup A_i(P)):K\cap L \neq \emptyset}\,\,\sum_{J: J \in P, K \subseteq J} \ckjpdp,
\end{align}
for all $L \subseteq D_i$. 
Finally, we have the following conditions on message rates and server capacities
\begin{align}\label{eq:combine}
R_{i}&=\sum_{ P \in \Pi: i \in I(P)}\,\,\sum_{\Dv \in \Delta(P)} R_{i}(P,\Dv), \quad & i \in [n],
\\\label{eq:combine2}
C_{J}&\geq\sum_{ P \in \Pi: J \in P} C_{J}(P) , \quad & J \in N.
\end{align}
The achievable rate region is characterized by eliminating the variables $(C_{J}(P), P \in \Pi, J \in P)$, $(R_{i}(P,\Dv), P \in \Pi, i \in I(P), \Dv \in \Delta(P))$, and $(\ckjpdp, P\in \Pi, J \in P, K \subseteq J, \Dv \in \Delta(P))$. 

A possible advantage of this technique is that each receiver only recovers composite indices of the groups that hold its desired message.
Consider the index coding problem
$$(1|2,3),\, (2|1,3),\, (3|1,2),\,(4|-).$$
Without server grouping, for receiver $i=4$ and server $J = \{1,2\}$ we have
\begin{align*}
 &\sum_{\Dv} \sum_{K:K \nsubseteq A_{4}} \ckjd =
\sum_{\Dv} \big[S_{\{1\},\{1,2\}}(\Dv)\\&\quad\quad+S_{\{2\},\{1,2\}}(\Dv)+ S_{\{1,2\},\{1,2\}}(\Dv)\big]\le C_J.
  \end{align*}
With partitioning, let us consider the server group $P = \{\{1,2\},\{2,3\}\}$. Since $i = 4$ does not belong to this group, we  do not have the above constraint for that group. 
A disadvantage of this technique is its computational complexity.
For $n$ messages, the number of possible non-empty server subsets is $|\Pi| = 2^{2^n-1}-1$, which is doubly exponential in $n$. As a result, Fourier-Motzkin elimination is not practical. Linear programs can be solved by excluding some servers from the problem (for example through dealing with trivial singleton servers, $|J| = 1$, in $N$ separately), by considering only a subset of $\Pi$ (at the possible expense of some reduction in the rate region), or when $\max |\Delta|$ is not very large.
Clearly the fractional distributed composite coding scheme reduces to the all-server scheme of Section \ref{sec:distributed:bounds:all:servers} when all the servers are put into a single group. At this point, however, we do not know any example for which the inner bound corresponding to the fractional distributed composite coding is strictly larger.

\section{Outer Bounds}\label{sec:outer:bounds}

\newtheorem{theo}{Theorem}
\newtheorem{lem}{Lemma}
\newtheorem{defi}{Definition}
\newtheorem{condi}{Condition}
\newtheorem{exam}{Example}
\newtheorem{remar}{Remark}

\setlength{\tabcolsep}{5pt}\definecolor{light-gray}{gray}{0.9}\renewcommand{\opertype}[1]{\begin{minipage}{145mm}\centering\vspace{1mm} #1\vspace{1mm}\end{minipage}}\rowcolors{1}{light-gray}{white}\footnotesize\begin{table*}[thb]\label{tab:results}
\begin{center}
\caption{Achievable sum-rate of all distributed index coding problems with $n=4$ in increasing order. Except for open problems $81, 112, 115, 119,$ and $148$, the achievable sum-rate matches the outer bound, thus establishing the sum-capacity.}
\begin{tabular}{|c|c|}\myhline $R_1+R_2+R_3+R_4$&Problem No\\\myhline 15&\opertype{1, 2, 3, 5, 6, 7, 8, 10, 11, 12, 13, 15, 17, 19, 20, 22, 25, 26, 33, 35, 38, 39, 40, 41, 49, 63, 65, 67, 69, 70, 100}\\\myhline18.667&\opertype{47}\\\myhline19&\opertype{\textbf{4, 9, 18, 21, 23, 24, 34, 36, 48, 55, 64, 66, 68, 86, 95, 99, 138}, $\overline{16, 30, 60, 102}$}\\\myhline20&\opertype{43, 78, 83, 85, 130, 132}\\\myhline21&\opertype{\textbf{14, 27, 28, 29, 31, 32, 37, 50, 51, 52, 53, 54, 56, 57, 58, 59, 61, 62, 87, 88, 89, 90, 91, 92, 94, 96, 97, 98, 101, 134, 136, 137, 139, 140, 141, 173}}\\\myhline22&\opertype{42, 44, 45, 71, 72, 73, 74, 75, 76, 77, 79, 80, 82, 84, 103, 104, 105, 106, 107, 108, 109, 110, 111, 113, 116, 117, 118, 120, 122, 123, 124, 125, 126, 127, 128, 131, 133, 142, 143, 144, 145, 147, 151, 152, 153, 154, 158, 159, 161, 162, 163, 164, 165, 166, 167, 168, 169, 174, 177, 182, 183, 184, 185, 186, 187, 201}\\\myhline23.333&\opertype{$\overline{46}$}\\\myhline23.5&\opertype{$81^{*}, 112^{*}, 115^{*}, 119^{*}, 148^{*}$}\\\myhline24&\opertype{114, 121, 129, 146, 150, 155, 156, 157, 160, 170, 171, 175, 178, 180, 181, 188, 189, 190, 191, 192, 194, 195, 196, 197, 198, 202, 204, 206, 208, 210, 216}\\\myhline25&\opertype{\underline{93, 135, 172, 199}}\\\myhline26&\opertype{207, \underline{\underline{149, 176, 179, 200, 203, 212}}}\\\myhline28&\opertype{193, 205, 209, 211, 213, 214, 215, 217}\\\myhline32&\opertype{218}\\\myhline\end{tabular}\end{center}\vspace{-0.5cm}\end{table*}

\normalsize

In this section, we use the shorthand notation $X_{[n]}$ to represent all messages $(X_{i},i\in [n])$ and $Y_N$ to represent all server outputs $(Y_{J}, J\in N)$. For any set $U$, we use $\overline U$ to represent the complimentary set of $U$ (when no confusion about the ground set will arise), and for any two sets $U$ and $V$, we use $UV$ to represent $U \cup V$. Note that due to source independence, $H(X_U|X_V) = H(X_U)$ for any two disjoint sets $U, V \subseteq [n]$. Also, the exact decoding condition at receiver $i$ stipulates $H(X_i|Y_N,X(A_i))=0$. We note that an $\epsilon$-error probability can be handled in a standard manner through converse techniques that use Fano's inequality. Finally, note that sever outputs only depend on the messages they contain. For example, $H(Y_N|X_{[n]}) = 0$. 

We now briefly review an outer bound on the capacity region from \cite{Sadeghi--Arbabjolfaei--Kim2016}, which is based on the polymatroidal axioms. 

\begin{theo}\label{theo:DPM}
Let $B_i$ be the set of interfering messages at receiver $i$, i.e., $B_i=[n]\backslash(A_i \cup \{i\})$. If $(\mathbf R, \mathbf C)$ is achievable, then for every $T \in N$ and every $i \in T$,
\begin{align}
R_i \leq f_T(B_i \cup \{i\})-f_T(B_i)
\end{align}
for some $f_T(S), S \subseteq T$ satisfying the polymatroidal axioms.
\end{theo}

The above theorem does not result in tight outer bounds in general, as we will show through numerical examples later. Now we introduce a new outer bound on the sum-rate based on Shannon inequalities, which can be strictly tighter. For simplicity of exposition, we focus on $\sum_{i=1}^{n}R_i$, but a subset of rates can also be considered in the sum. The following two disjoint sets $U,V$ will prove useful.




\begin{defi}\label{def:set:C}
Given a distributed index coding problem, $U$ is the largest set such that $U \subseteq [n]$ and $H(X_U|Y_N)=0$.
\end{defi}

\begin{defi}\label{def:set:B}
Given a distributed index coding problem and its corresponding set $U$ from Definition \ref{def:set:C}, set $V$ is the smallest set that satisfies $V\subseteq [n] \backslash U$ and $H(X_{\overline{UV}}|Y_N,X_{UV})=0$.
\end{defi}

Note that set $U$ is unique and can be found through exhaustive search. However, for some index coding problems, there can be multiple sets $V$.

\begin{exam}\label{exam:140:1}
Consider the problem $(1|-)$; $(2|1, 4)$; $(3|1, 2)$; $(4|1, 2, 3)$. Using $H(X_i|Y_N,X(A_i))=0, i \in [n]$, we find that $U = \{1\}$ and $V = \{2\}$ or $V =\{4\}$.
\end{exam}

Lemma \ref{lem:k} below comes directly from Definitions \ref{def:set:C} and \ref{def:set:B}.
\begin{lem}\label{lem:k}
Given an index coding problem and sets $U$ and $V$ defined in Definitions \ref{def:set:C} and \ref{def:set:B}, we always have
\begin{align}
H(X_{[n]})&=H(X_{[n]},Y_N)=H(Y_N)+H(X_{[n]}|Y_N)\\\nonumber
                             &=H(Y_N)+H(X_U |Y_N)+H(X_V|Y_N,X_U)\\
                             &\quad\quad\quad+H(X_{\overline{UV}}|Y_N,X_{UV})\\
                             &=H(Y_N)+H(X_V|Y_N,X_U)
\end{align}
\end{lem}
Now we consider the following condition that is used to establish a new outer bound on the sum-rate.

\begin{condi}\label{cond:condientro:zero}
For any set $V \subseteq [n]$, if $H(X_V|Y_N,X_{\overline V})=0$, then we say that set $V$ satisfies Condition \ref{cond:condientro:zero}.
\end{condi}

\begin{theo}\label{theo:BC}
Given a distributed index coding problem, if set $V$ from Definition \ref{def:set:B} satisfies Condition \ref{cond:condientro:zero}, then we have the following outer bound on the sum-rate:
\begin{align}
\sum\limits_{i\in [n]}R_i\leq \sum\limits_{J\in N}C_J+\sum\limits_{J\in N:J\cap V\neq \emptyset, J\nsubseteq UV}C_J
\end{align}
\end{theo}

\begin{IEEEproof}
Consider $H(X_V)$ and $I(X_V;Y_{J:J\cap V\neq \emptyset}|X_U)$:
\begin{align}
H(X_V)&=H(X_V|X_{\overline{V}}) \label{eq:source:inde} 
            =H(X_V,Y_{J:J\cap V\neq \emptyset}|X_{\overline{V}})\\
            &=H(Y_{J:J\cap V\neq \emptyset}|X_{\overline{V}})+H(X_V|X_{\overline{V}},Y_N)\\
            &=H(Y_{J:J\cap V\neq \emptyset}|X_{\overline{V}}) \label{eq:condientro:zero}
\end{align}
where \eqref{eq:condientro:zero} is due to the fact that set $V$ satisfies Condition \ref{cond:condientro:zero}.
\begin{align}
&I(X_V;Y_{J:J\cap V\neq \emptyset}|X_U)\\&=H(X_V)-H(X_V|Y_{J:J\cap V\neq \emptyset},X_U) \label{eq:i:1} \\
                                                                            &=H(Y_{J:J\cap V\neq \emptyset}|X_U)-H(Y_{J:J\cap V\neq \emptyset,J\nsubseteq UV}|X_{UV}) \label{eq:i:2}
\end{align}
Since $H(X_V)=H(Y_{J:J\cap V\neq \emptyset}|X_{\overline{V}})\leq H(Y_{J:J\cap V\neq \emptyset}|X_U)$ and $H(X_V|Y_N,X_U)\leq H(X_V|{Y_{J:J\cap V\neq \emptyset}},X_U)$, by comparing (\ref{eq:i:1}) and (\ref{eq:i:2}), we have
\begin{align}
H(X_V|Y_N,X_U) \leq H(Y_{J:J\cap V\neq \emptyset,J\nsubseteq UV}|X_{UV})
\end{align}
Therefore, we have
\begin{align}
\sum\limits_{i\in [n]}R_i &\leq H(X_{[n]})
                                      =H(Y_N)+H(X_V|Y_N,X_U) \label{eq:lem:k} \\
                                      &\leq H(Y_N)+H(Y_{J:J\cap V\neq \emptyset,J\nsubseteq UV}|X_{UV})\\
                                     &\leq \sum\limits_{J\in N}C_J+\sum\limits_{J\in N:J\cap V\neq \emptyset, J\nsubseteq UV}C_J
\end{align}
where the equality in (\ref{eq:lem:k}) follows from Lemma \ref{lem:k}.
\end{IEEEproof}

\begin{exam}\label{example:140:2} Revisiting Example \ref{exam:140:1} and directly applying Theorem \ref{theo:BC}, for $V = \{2\}$ we will have
\begin{align}
\sum\limits_{i\in [n]}R_i &\leq \sum_{J\in N}C_J+\sum_{J\in N:J\cap \{2\} \neq \emptyset, J\nsubseteq \{1,2\}}C_J 
\end{align}
And for $V = \{4\}$ we will have
\begin{align}
\sum\limits_{i\in [n]}R_i &\leq \sum_{J\in N}C_J+\sum_{J\in N:J\cap \{4\} \neq \emptyset, J\nsubseteq \{1,4\}}C_J 
\end{align}
If $C_J=1, \forall J$, then both sets $V=\{2\}$ and $V = \{4\}$ give the same outer bound of 21 on the sum-rate. However, in general one needs to evaluate all bounds  due to different sets $V$ and take the minimum. Note that with $C_J=1, \forall J$, Theorem \ref{theo:DPM} gives $\sum\limits_{i\in [n]}R_i \leq 22$, which is strictly looser.





\end{exam}



\section{Numerical Results}\label{sec:numerical}

We numerically evaluated inner and outer bounds on the sum-rate of all 218 non-isomorphic distributed index coding problems with $n=4$ messages. The receiver side information of these problems are listed in the next section. For instance, the side information in Examples \ref{exam:140:1} and \ref{example:140:2} is labelled as problem number 140. The inner bound is computed by applying LP on the composite coding scheme of Section \ref{sec:distributed:bounds:all:servers} and compares with the outer bound on the sum-rate as follows.

\begin{itemize}
\item For 145 out of 218 problems, the outer bound on the sum-rate due to Theorem  \ref{theo:DPM} and the inner bound matched. These cases are shown normally in Table \ref{tab:results}.
\item For 53 out of the remaining 73 cases, the outer bound of Theorem \ref{theo:BC} gave a tighter result than Theorem  \ref{theo:DPM} and matched the inner bound. These cases are shown in bold font in Table \ref{tab:results}. For the final 20 cases, the outer bounds of Theorems \ref{theo:DPM} or \ref{theo:BC} did not match the inner bound.
\item For 4 of these final 20 cases, there did not exist a single set $V$ that satisfied Condition \ref{cond:condientro:zero}. However, we partitioned $V$ into several disjoint subsets such that each part satisfied Condition \ref{cond:condientro:zero}. Then we used a method similar to Theorem \ref{theo:BC} and obtained a tighter outer bound that matched the inner bound. These cases are shown as underlined in Table \ref{tab:results}.
\item For 6 of the final 20 cases, we had to use non-trivial extensions of Theorem \ref{theo:BC}, not presented in this paper due to space limitations. The obtained tighter outer bound  matched the inner bound. These cases are shown as doubly underlined in Table \ref{tab:results}.
\item For 5 of the remaining problems, we had to use f-d separation methods based on \cite{satyajit, kramer} to obtain a tighter outer bound that matched the inner bound. The details are not presented in this paper due to space limitations. These cases are shown as overlined in Table \ref{tab:results}. The sum-capacity of five problems 81, 112, 115, 119, and 148 remains open. The tightest outer bound on the sum-rate that we have achieved is $23 \frac{2}{3}$, which we conjecture can be further tightened to $23.5$.
\end{itemize}

To conclude, we report that for 28 out of these 218  problems, the composite coding scheme of \cite[Section IV-B]{Sadeghi--Arbabjolfaei--Kim2016} gives a looser inner bound on the sum-rate than the enhanced method of Section \ref{sec:distributed:bounds:all:servers} in this paper. For example, for problem number 155: $(1|4);(2|3,4);(3|1,2);(4|2,3)$, \cite[Section IV-B]{Sadeghi--Arbabjolfaei--Kim2016} gives $\sum_{i \in [n]} R_i \leq 23$, whereas the sum-capacity 24 is achievable through  the method of Section \ref{sec:distributed:bounds:all:servers}. In effect, server partitioning of \cite[Section IV-C]{Sadeghi--Arbabjolfaei--Kim2016} is not necessary to achieve the sum-capacity of problem no 155. It remains unclear  whether and when the fractional method of Section \ref{sec:distributed:bounds:fractional} offers strict improvement over the all-server method of Section \ref{sec:distributed:bounds:all:servers}. This is our ongoing research.

\bibliographystyle{IEEEtran}
\bibliography{references} 

\section{List of all Non-isomorphic Index Coding Problems with $n=4$ Messages}
\noindent Problem No 1: $(1|-),(2|-),(3|-),(4|-)$\\ 
Problem No 2: $(1|2),(2|-),(3|-),(4|-)$\\ 
Problem No 3: $(1|2,3),(2|-),(3|-),(4|-)$\\ 
Problem No 4: $(1|-),(2|-),(3|4),(4|3)$\\ 
Problem No 5: $(1|-),(2|-),(3|4),(4|2)$\\ 
Problem No 6: $(1|-),(2|-),(3|2),(4|2)$\\ 
Problem No 7: $(1|-),(2|-),(3|2),(4|1)$\\ 
Problem No 8: $(1|2,3,4),(2|-),(3|-),(4|-)$\\ 
Problem No 9: $(1|-),(2|-),(3|4),(4|2,3)$\\ 
Problem No 10: $(1|-),(2|-),(3|4),(4|1,2)$\\ 
Problem No 11: $(1|-),(2|-),(3|2),(4|2,3)$\\ 
Problem No 12: $(1|-),(2|-),(3|2),(4|1,3)$\\ 
Problem No 13: $(1|-),(2|-),(3|2),(4|1,2)$\\ 
Problem No 14: $(1|-),(2|4),(3|4),(4|3)$\\ 
Problem No 15: $(1|-),(2|4),(3|4),(4|1)$\\ 
Problem No 16: $(1|-),(2|4),(3|2),(4|3)$\\ 
Problem No 17: $(1|-),(2|4),(3|2),(4|1)$\\ 
Problem No 18: $(1|-),(2|4),(3|1),(4|2)$\\ 
Problem No 19: $(1|-),(2|4),(3|1),(4|1)$\\ 
Problem No 20: $(1|-),(2|1),(3|1),(4|1)$\\ 
Problem No 21: $(1|2,3,4),(2|1),(3|-),(4|-)$\\ 
Problem No 22: $(1|-),(2|-),(3|2),(4|1,2,3)$\\ 
Problem No 23: $(1|-),(2|-),(3|2,4),(4|2,3)$\\ 
Problem No 24: $(1|-),(2|-),(3|2,4),(4|1,3)$\\ 
Problem No 25: $(1|-),(2|-),(3|2,4),(4|1,2)$\\ 
Problem No 26: $(1|-),(2|-),(3|1,2),(4|1,2)$\\ 
Problem No 27: $(1|-),(2|4),(3|4),(4|2,3)$\\ 
Problem No 28: $(1|-),(2|4),(3|4),(4|1,3)$\\ 
Problem No 29: $(1|-),(2|4),(3|2),(4|2,3)$\\ 
Problem No 30: $(1|-),(2|4),(3|2),(4|1,3)$\\ 
Problem No 31: $(1|-),(2|4),(3|2),(4|1,2)$\\ 
Problem No 32: $(1|-),(2|4),(3|2,4),(4|2)$\\ 
Problem No 33: $(1|-),(2|4),(3|2,4),(4|1)$\\ 
Problem No 34: $(1|-),(2|4),(3|1),(4|2,3)$\\ 
Problem No 35: $(1|-),(2|4),(3|1),(4|1,3)$\\ 
Problem No 36: $(1|-),(2|4),(3|1),(4|1,2)$\\ 
Problem No 37: $(1|-),(2|4),(3|1,4),(4|2)$\\ 
Problem No 38: $(1|-),(2|4),(3|1,4),(4|1)$\\ 
Problem No 39: $(1|-),(2|4),(3|1,2),(4|1)$\\ 
Problem No 40: $(1|-),(2|3,4),(3|1),(4|1)$\\ 
Problem No 41: $(1|-),(2|1),(3|1),(4|1,3)$\\ 
Problem No 42: $(1|4),(2|4),(3|4),(4|3)$\\ 
Problem No 43: $(1|4),(2|4),(3|2),(4|3)$\\ 
Problem No 44: $(1|4),(2|4),(3|2),(4|2)$\\ 
Problem No 45: $(1|4),(2|4),(3|2),(4|1)$\\ 
Problem No 46: $(1|4),(2|3),(3|2),(4|1)$\\ 
Problem No 47: $(1|4),(2|3),(3|1),(4|2)$\\ 
Problem No 48: $(1|2,3,4),(2|1,3),(3|-),(4|-)$\\ 
Problem No 49: $(1|-),(2|-),(3|1,2),(4|1,2,3)$\\ 
Problem No 50: $(1|-),(2|4),(3|4),(4|1,2,3)$\\ 
Problem No 51: $(1|-),(2|4),(3|2),(4|1,2,3)$\\ 
Problem No 52: $(1|-),(2|4),(3|2,4),(4|2,3)$\\ 
Problem No 53: $(1|-),(2|4),(3|2,4),(4|1,3)$\\ 
Problem No 54: $(1|-),(2|4),(3|2,4),(4|1,2)$\\ 
Problem No 55: $(1|-),(2|4),(3|1),(4|1,2,3)$\\ 
Problem No 56: $(1|-),(2|4),(3|1,4),(4|2,3)$\\ 
Problem No 57: $(1|-),(2|4),(3|1,4),(4|1,3)$\\ 
Problem No 58: $(1|-),(2|4),(3|1,4),(4|1,2)$\\ 
Problem No 59: $(1|-),(2|4),(3|1,2),(4|2,3)$\\ 
Problem No 60: $(1|-),(2|4),(3|1,2),(4|1,3)$\\ 
Problem No 61: $(1|-),(2|4),(3|1,2),(4|1,2)$\\ 
Problem No 62: $(1|-),(2|4),(3|1,2,4),(4|2)$\\ 
Problem No 63: $(1|-),(2|4),(3|1,2,4),(4|1)$\\ 
Problem No 64: $(1|-),(2|3,4),(3|2,4),(4|1)$\\ 
Problem No 65: $(1|-),(2|3,4),(3|1),(4|1,3)$\\ 
Problem No 66: $(1|-),(2|3,4),(3|1),(4|1,2)$\\ 
Problem No 67: $(1|-),(2|1),(3|1),(4|1,2,3)$\\ 
Problem No 68: $(1|-),(2|1),(3|1,4),(4|1,3)$\\ 
Problem No 69: $(1|-),(2|1),(3|1,4),(4|1,2)$\\ 
Problem No 70: $(1|-),(2|1),(3|1,2),(4|1,2)$\\ 
Problem No 71: $(1|4),(2|4),(3|4),(4|2,3)$\\ 
Problem No 72: $(1|4),(2|4),(3|2),(4|2,3)$\\ 
Problem No 73: $(1|4),(2|4),(3|2),(4|1,3)$\\ 
Problem No 74: $(1|4),(2|4),(3|2),(4|1,2)$\\ 
Problem No 75: $(1|4),(2|4),(3|2,4),(4|3)$\\ 
Problem No 76: $(1|4),(2|4),(3|2,4),(4|2)$\\ 
Problem No 77: $(1|4),(2|4),(3|2,4),(4|1)$\\ 
Problem No 78: $(1|4),(2|4),(3|1,2),(4|3)$\\ 
Problem No 79: $(1|4),(2|4),(3|1,2),(4|2)$\\ 
Problem No 80: $(1|4),(2|3),(3|2),(4|2,3)$\\ 
Problem No 81: $(1|4),(2|3),(3|2),(4|1,3)$\\ 
Problem No 82: $(1|4),(2|3),(3|2,4),(4|2)$\\ 
Problem No 83: $(1|4),(2|3),(3|1),(4|2,3)$\\ 
Problem No 84: $(1|4),(2|3),(3|1),(4|1,2)$\\ 
Problem No 85: $(1|4),(2|3,4),(3|1),(4|3)$\\ 
Problem No 86: $(1|2,3,4),(2|1,3,4),(3|-),(4|-)$\\ 
Problem No 87: $(1|-),(2|4),(3|2,4),(4|1,2,3)$\\ 
Problem No 88: $(1|-),(2|4),(3|1,4),(4|1,2,3)$\\ 
Problem No 89: $(1|-),(2|4),(3|1,2),(4|1,2,3)$\\ 
Problem No 90: $(1|-),(2|4),(3|1,2,4),(4|2,3)$\\ 
Problem No 91: $(1|-),(2|4),(3|1,2,4),(4|1,3)$\\ 
Problem No 92: $(1|-),(2|4),(3|1,2,4),(4|1,2)$\\ 
Problem No 93: $(1|-),(2|3,4),(3|2,4),(4|2,3)$\\ 
Problem No 94: $(1|-),(2|3,4),(3|2,4),(4|1,3)$\\ 
Problem No 95: $(1|-),(2|3,4),(3|1),(4|1,2,3)$\\ 
Problem No 96: $(1|-),(2|3,4),(3|1,4),(4|1,3)$\\ 
Problem No 97: $(1|-),(2|3,4),(3|1,4),(4|1,2)$\\ 
Problem No 98: $(1|-),(2|3,4),(3|1,2),(4|1,2)$\\ 
Problem No 99: $(1|-),(2|1),(3|1,4),(4|1,2,3)$\\ 
Problem No 100: $(1|-),(2|1),(3|1,2),(4|1,2,3)$\\ 
Problem No 101: $(1|-),(2|1,4),(3|1,4),(4|1,3)$\\ 
Problem No 102: $(1|-),(2|1,4),(3|1,2),(4|1,3)$\\ 
Problem No 103: $(1|4),(2|4),(3|4),(4|1,2,3)$\\ 
Problem No 104: $(1|4),(2|4),(3|2),(4|1,2,3)$\\ 
Problem No 105: $(1|4),(2|4),(3|2,4),(4|2,3)$\\ 
Problem No 106: $(1|4),(2|4),(3|2,4),(4|1,3)$\\ 
Problem No 107: $(1|4),(2|4),(3|2,4),(4|1,2)$\\ 
Problem No 108: $(1|4),(2|4),(3|1,2),(4|2,3)$\\ 
Problem No 109: $(1|4),(2|4),(3|1,2),(4|1,2)$\\ 
Problem No 110: $(1|4),(2|4),(3|1,2,4),(4|3)$\\ 
Problem No 111: $(1|4),(2|4),(3|1,2,4),(4|2)$\\ 
Problem No 112: $(1|4),(2|3),(3|2),(4|1,2,3)$\\ 
Problem No 113: $(1|4),(2|3),(3|2,4),(4|2,3)$\\ 
Problem No 114: $(1|4),(2|3),(3|2,4),(4|1,3)$\\ 
Problem No 115: $(1|4),(2|3),(3|2,4),(4|1,2)$\\ 
Problem No 116: $(1|4),(2|3),(3|1),(4|1,2,3)$\\ 
Problem No 117: $(1|4),(2|3),(3|1,4),(4|2,3)$\\ 
Problem No 118: $(1|4),(2|3),(3|1,4),(4|1,2)$\\ 
Problem No 119: $(1|4),(2|3),(3|1,2),(4|1,2)$\\ 
Problem No 120: $(1|4),(2|3,4),(3|2,4),(4|3)$\\ 
Problem No 121: $(1|4),(2|3,4),(3|2,4),(4|1)$\\ 
Problem No 122: $(1|4),(2|3,4),(3|1),(4|2,3)$\\ 
Problem No 123: $(1|4),(2|3,4),(3|1),(4|1,3)$\\ 
Problem No 124: $(1|4),(2|3,4),(3|1),(4|1,2)$\\ 
Problem No 125: $(1|4),(2|3,4),(3|1,4),(4|3)$\\ 
Problem No 126: $(1|4),(2|3,4),(3|1,4),(4|2)$\\ 
Problem No 127: $(1|4),(2|3,4),(3|1,4),(4|1)$\\ 
Problem No 128: $(1|4),(2|3,4),(3|1,2),(4|3)$\\ 
Problem No 129: $(1|4),(2|3,4),(3|1,2),(4|1)$\\ 
Problem No 130: $(1|4),(2|1),(3|1,2),(4|2,3)$\\ 
Problem No 131: $(1|4),(2|1),(3|1,2),(4|1,2)$\\ 
Problem No 132: $(1|4),(2|1),(3|1,2,4),(4|2)$\\ 
Problem No 133: $(1|4),(2|1,4),(3|1,4),(4|1)$\\ 
Problem No 134: $(1|2,3,4),(2|1,3,4),(3|1),(4|-)$\\ 
Problem No 135: $(1|-),(2|3,4),(3|2,4),(4|1,2,3)$\\ 
Problem No 136: $(1|-),(2|3,4),(3|1,4),(4|1,2,3)$\\ 
Problem No 137: $(1|-),(2|3,4),(3|1,2),(4|1,2,3)$\\ 
Problem No 138: $(1|-),(2|1),(3|1,2,4),(4|1,2,3)$\\ 
Problem No 139: $(1|-),(2|1,4),(3|1,4),(4|1,2,3)$\\ 
Problem No 140: $(1|-),(2|1,4),(3|1,2),(4|1,2,3)$\\ 
Problem No 141: $(1|-),(2|1,4),(3|1,2,4),(4|1,2)$\\ 
Problem No 142: $(1|4),(2|4),(3|2,4),(4|1,2,3)$\\ 
Problem No 143: $(1|4),(2|4),(3|1,2),(4|1,2,3)$\\ 
Problem No 144: $(1|4),(2|4),(3|1,2,4),(4|2,3)$\\ 
Problem No 145: $(1|4),(2|4),(3|1,2,4),(4|1,2)$\\ 
Problem No 146: $(1|4),(2|3),(3|2,4),(4|1,2,3)$\\ 
Problem No 147: $(1|4),(2|3),(3|1,4),(4|1,2,3)$\\ 
Problem No 148: $(1|4),(2|3),(3|1,2),(4|1,2,3)$\\ 
Problem No 149: $(1|4),(2|3,4),(3|2,4),(4|2,3)$\\ 
Problem No 150: $(1|4),(2|3,4),(3|2,4),(4|1,3)$\\ 
Problem No 151: $(1|4),(2|3,4),(3|1),(4|1,2,3)$\\ 
Problem No 152: $(1|4),(2|3,4),(3|1,4),(4|2,3)$\\ 
Problem No 153: $(1|4),(2|3,4),(3|1,4),(4|1,3)$\\ 
Problem No 154: $(1|4),(2|3,4),(3|1,4),(4|1,2)$\\ 
Problem No 155: $(1|4),(2|3,4),(3|1,2),(4|2,3)$\\ 
Problem No 156: $(1|4),(2|3,4),(3|1,2),(4|1,3)$\\ 
Problem No 157: $(1|4),(2|3,4),(3|1,2),(4|1,2)$\\ 
Problem No 158: $(1|4),(2|3,4),(3|1,2,4),(4|3)$\\ 
Problem No 159: $(1|4),(2|3,4),(3|1,2,4),(4|2)$\\ 
Problem No 160: $(1|4),(2|3,4),(3|1,2,4),(4|1)$\\ 
Problem No 161: $(1|4),(2|1),(3|1,2),(4|1,2,3)$\\ 
Problem No 162: $(1|4),(2|1),(3|1,2,4),(4|2,3)$\\ 
Problem No 163: $(1|4),(2|1),(3|1,2,4),(4|1,2)$\\ 
Problem No 164: $(1|4),(2|1,4),(3|1,4),(4|2,3)$\\ 
Problem No 165: $(1|4),(2|1,4),(3|1,4),(4|1,3)$\\ 
Problem No 166: $(1|4),(2|1,4),(3|1,2),(4|2,3)$\\ 
Problem No 167: $(1|4),(2|1,4),(3|1,2),(4|1,3)$\\ 
Problem No 168: $(1|4),(2|1,4),(3|1,2),(4|1,2)$\\ 
Problem No 169: $(1|4),(2|1,4),(3|1,2,4),(4|1)$\\ 
Problem No 170: $(1|4),(2|1,3),(3|1,2),(4|2,3)$\\ 
Problem No 171: $(1|4),(2|1,3),(3|1,2),(4|1,3)$\\ 
Problem No 172: $(1|2,3,4),(2|1,3,4),(3|1,2),(4|-)$\\ 
Problem No 173: $(1|-),(2|1,4),(3|1,2,4),(4|1,2,3)$\\ 
Problem No 174: $(1|4),(2|4),(3|1,2,4),(4|1,2,3)$\\ 
Problem No 175: $(1|4),(2|3),(3|1,2,4),(4|1,2,3)$\\ 
Problem No 176: $(1|4),(2|3,4),(3|2,4),(4|1,2,3)$\\ 
Problem No 177: $(1|4),(2|3,4),(3|1,4),(4|1,2,3)$\\ 
Problem No 178: $(1|4),(2|3,4),(3|1,2),(4|1,2,3)$\\ 
Problem No 179: $(1|4),(2|3,4),(3|1,2,4),(4|2,3)$\\ 
Problem No 180: $(1|4),(2|3,4),(3|1,2,4),(4|1,3)$\\ 
Problem No 181: $(1|4),(2|3,4),(3|1,2,4),(4|1,2)$\\ 
Problem No 182: $(1|4),(2|1),(3|1,2,4),(4|1,2,3)$\\ 
Problem No 183: $(1|4),(2|1,4),(3|1,4),(4|1,2,3)$\\ 
Problem No 184: $(1|4),(2|1,4),(3|1,2),(4|1,2,3)$\\ 
Problem No 185: $(1|4),(2|1,4),(3|1,2,4),(4|2,3)$\\ 
Problem No 186: $(1|4),(2|1,4),(3|1,2,4),(4|1,3)$\\ 
Problem No 187: $(1|4),(2|1,4),(3|1,2,4),(4|1,2)$\\ 
Problem No 188: $(1|4),(2|1,3),(3|1,2),(4|1,2,3)$\\ 
Problem No 189: $(1|4),(2|1,3),(3|1,2,4),(4|2,3)$\\ 
Problem No 190: $(1|4),(2|1,3),(3|1,2,4),(4|1,3)$\\ 
Problem No 191: $(1|4),(2|1,3),(3|1,2,4),(4|1,2)$\\ 
Problem No 192: $(1|4),(2|1,3,4),(3|1,2,4),(4|1)$\\ 
Problem No 193: $(1|3,4),(2|3,4),(3|2,4),(4|2,3)$\\ 
Problem No 194: $(1|3,4),(2|3,4),(3|2,4),(4|1,3)$\\ 
Problem No 195: $(1|3,4),(2|3,4),(3|2,4),(4|1,2)$\\ 
Problem No 196: $(1|3,4),(2|3,4),(3|1,2),(4|1,2)$\\ 
Problem No 197: $(1|3,4),(2|1,4),(3|2,4),(4|2,3)$\\ 
Problem No 198: $(1|3,4),(2|1,4),(3|1,2),(4|2,3)$\\ 
Problem No 199: $(1|2,3,4),(2|1,3,4),(3|1,2,4),(4|-)$\\ 
Problem No 200: $(1|4),(2|3,4),(3|1,2,4),(4|1,2,3)$\\ 
Problem No 201: $(1|4),(2|1,4),(3|1,2,4),(4|1,2,3)$\\ 
Problem No 202: $(1|4),(2|1,3),(3|1,2,4),(4|1,2,3)$\\ 
Problem No 203: $(1|4),(2|1,3,4),(3|1,2,4),(4|2,3)$\\ 
Problem No 204: $(1|4),(2|1,3,4),(3|1,2,4),(4|1,3)$\\ 
Problem No 205: $(1|3,4),(2|3,4),(3|2,4),(4|1,2,3)$\\ 
Problem No 206: $(1|3,4),(2|3,4),(3|1,2),(4|1,2,3)$\\ 
Problem No 207: $(1|3,4),(2|1,4),(3|2,4),(4|1,2,3)$\\ 
Problem No 208: $(1|3,4),(2|1,4),(3|1,2),(4|1,2,3)$\\ 
Problem No 209: $(1|3,4),(2|1,4),(3|1,2,4),(4|1,3)$\\ 
Problem No 210: $(1|3,4),(2|1,4),(3|1,2,4),(4|1,2)$\\ 
Problem No 211: $(1|3,4),(2|1,3,4),(3|1,4),(4|1,3)$\\ 
Problem No 212: $(1|2,3,4),(2|1,3,4),(3|1,2,4),(4|1)$\\ 
Problem No 213: $(1|3,4),(2|3,4),(3|1,2,4),(4|1,2,3)$\\ 
Problem No 214: $(1|3,4),(2|1,4),(3|1,2,4),(4|1,2,3)$\\ 
Problem No 215: $(1|3,4),(2|1,3,4),(3|1,4),(4|1,2,3)$\\ 
Problem No 216: $(1|3,4),(2|1,3,4),(3|1,2),(4|1,2,3)$\\ 
Problem No 217: $(1|2,3,4),(2|1,3,4),(3|1,2,4),(4|1,2)$\\ 
Problem No 218: $(1|2,3,4),(2|1,3,4),(3|1,2,4),(4|1,2,3)$\\ 





\end{document}